\documentclass[aps,prl,twocolumn,groupedaddress,showpacs]{revtex4}

\usepackage{graphicx}
\usepackage{dcolumn}
\usepackage{bm}
\usepackage{subfigure}

\begin{document}

\preprint{}

\title{Direct excitation of the forbidden clock transition in neutral $^{174}$Yb atoms confined to an optical lattice}

\author{Z. W. Barber}
\altaffiliation{University of Colorado, Boulder, CO, 80309}
\email{zbarber@boulder.nist.gov}
\author{C. W. Hoyt}
\author{C. W. Oates}
\author{L. Hollberg}
\affiliation{National Institute of Standards and Technology\\325
Broadway, Boulder, CO 80305}
\thanks{Official contribution of the National Institute of Standards and Technology; not subject to copyright.}
\author{A. V. Taichenachev}
\author{V. I. Yudin}
\affiliation{Institute of Laser Physics SB RAS, Novosibirsk 630090, Russia\\ and Novosibirsk State University, Novosibirsk 630090, Russia}
\date{\today}

\begin{abstract}
We report direct single-laser excitation of the strictly forbidden $(6s^2)\,^1\!S_0 \leftrightarrow (6s6p)\,^3\!P_0$ clock transition in $^{174}$Yb atoms confined to a 1D optical lattice. A small ($\sim$1.2\,mT) static magnetic field was used to induce a nonzero electric dipole transition probability between the clock states at 578.42\,nm.  Narrow resonance linewidths of 20\,Hz (FWHM) with high contrast were observed, demonstrating a resonance quality factor of $2.6\times10^{13}$.  The previously unknown ac Stark shift-canceling (magic) wavelength was determined to be $759.35\pm0.02$\,nm. This method for using the metrologically superior even isotope can be easily implemented in current Yb and Sr lattice clocks, and can create new clock possibilities in other alkaline earth-like atoms such as Mg and Ca.
\end{abstract}

\pacs{06.30.Ft, 32.80.Pj, 39.30.+w}

\maketitle

Time and frequency standards play an integral role in modern technologies and provide precise definitions of fundamental quantities. Optical frequency standards, which have distinct advantages relative to microwave frequency clocks, are rapidly approaching maturity \cite{Diddams04}. Considerable strides have been taken recently toward a new clock scheme based on large numbers of neutral atoms trapped in far off-resonance optical lattices \cite{Takamoto05,Ludlow05} that are designed to have zero net ac Stark shift on the clock transition. This type of clock promises to combine the high signal-to-noise ratios of current neutral atom standards \cite{Sterr04} with the long interaction times and Doppler-free spectroscopy found in single-ion standards \cite{Gill03}. In this Letter, we demonstrate the first spectroscopy of the Yb clock transition in a Stark-free lattice.  More importantly, this demonstration is the first precision spectroscopy of the $^1\!S_0 \leftrightarrow\,^3\!P_0$ clock transition in an even isotope of an alkaline earth-like atom.  Realizing a recently proposed method of magnetic field-induced spectroscopy \cite{Taichenachev05}, we observe spectroscopic linewidths of 20 Hz, full-width at half-maximum (FWHM), with good contrast and signal-to-noise ratios.  Yielding a resonance quality factor of $2.6\times10^{13}$, these results demonstrate the feasibility of a lattice clock based on the metrologically superior even isotope. 

Until now, lattice clock experiments have focused on the $^1\!S_0 \leftrightarrow\,^3\!P_0$ clock transition in the odd isotopes of alkaline earth-like atoms Sr \cite{Takamoto05,Ludlow05,Courtillot03} and Yb \cite{Porsev04,Park03,Hong05b,Hoyt05}.  This transition, weakly allowed in the odd isotopes due to hyperfine mixing, provides a narrow natural linewidth ($\sim$\,millihertz) and a high degree of lattice polarization insensitivity.  However, the nonzero nuclear spin of the odd isotopes creates undesirable residual lattice polarization sensitivity, optical pumping issues, and linear magnetic field sensitivity. The use of the even-isotope alkaline earth-like atoms (zero nuclear spin) would essentially eliminate these significant experimental difficulties.  Unfortunately, the direct optical excitation of this transition in the even isotopes is forbidden due to the absence of hyperfine mixing and has therefore, until now, remained out of reach.

Multiphoton schemes to excite this even-isotope transition in Yb and Sr have recently been proposed \cite{Santra05, Hong05}.  These proposals are based on relatively complex experimental apparatus, requiring the use of multiple stabilized lasers, nonlinear optics, and good control over laser field intensities.  In contrast, the demonstration described in this Letter uses only the addition of a small magnetic field to mix a fraction of a nearby state into the upper clock state, creating a weakly allowed electric dipole transition.  Thus, a single clock laser at 578.42\,nm is sufficient to directly excite the transition.  Although the bias field must be made stable, this method allows one to reap significant gains in simplicity.

With the inherent insensitivity of the even isotopes to lattice polarization and residual magnetic fields,
this approach has the potential to provide 17 digits of accuracy or more with only minor changes to existing lattice clock experiments.  In addition to Sr and Yb, the method can be employed with other group II atoms (e.g. Ca, Mg), thereby opening up new clock opportunities \cite{Taichenachev05}. This technique should expedite the development of high accuracy optical lattice clocks, enabling improved standards and aiding in the search for time variations of fundamental physical constants \cite{Karshenboim00}.

\begin{figure}
\includegraphics[width=8cm]{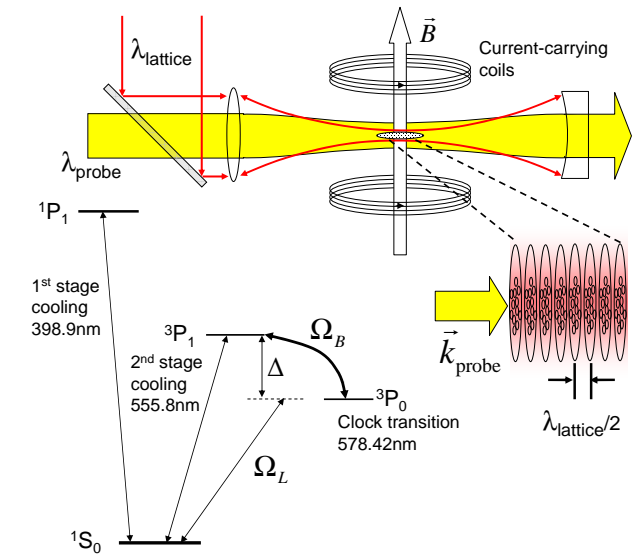}
\caption{Magnetic field-induced lattice spectroscopy.  Approximately 10\,000 $^{174}$Yb atoms were cooled, trapped,  and loaded into a lattice at the Stark-free (magic) wavelength of 759.35\,nm.  A pair of current-carrying coils generated a static magnetic field ($\Omega_B = \langle ^3\!P_1,m_J=0|\vec{\mu}\cdot{\bf B}|^3\!P_0\rangle/\hbar$) to mix a small portion of the $^3\!P_1$ state into the $^3\!P_0$ clock state split by a frequency $\Delta$.  A single 578.42\,nm spectroscopic probe laser beam, with Rabi frequency $\Omega_L$ on the $^1\!S_0\leftrightarrow\,^3\!P_1$ transition, was aligned collinear to the lattice with a dichroic beamsplitter.  The tight confinement of the atoms in the probe direction provided for Doppler- and recoil-free excitation of the clock transition. Excitation was measured by ground-state fluorescence of the atoms on the 398.9\,nm transition.}
\label{setup}
\end{figure}

This initial demonstration of magnetic-field induced spectroscopy was performed in a 1D optical lattice.  In order to capture atoms in the relatively weak trapping potential of the lattice, two transitions for laser cooling and trapping were utilized. Diode lasers tuned to the broad transition ($\Gamma/2\pi = 28$\,MHz) at 398.9\,nm were used for decelerating, cooling, and collecting Yb atoms in a magneto-optical trap (MOT). In 300\,ms of loading, roughly a million atoms were trapped with a temperature of a few millikelvin.  Then the violet light was extinguished and the gradient was reduced to capture up to $70\,\%$ of the atoms in a MOT based on the 555.8\,nm intercombination line ($\Gamma/2\pi=182$\,kHz). (See Ref. \cite{Hoyt05} for more details.)  After the atoms reached equilibrium in this green MOT ($\sim$\,25\,ms), the light intensity was turned down by a factor of 10 to 100 for 25\,ms to reduce the temperature of the atoms to roughly $50\,\mu$K for loading into the far off-resonance lattice, which was on continuously. Roughly $10^4$ atoms remained trapped in the Stark-free lattice (lifetime $\approx$ 150\,ms) after all the MOT light and magnetic gradients are turned off. A single spectroscopic pulse (with bias magnetic field turned on for even isotopes) then excited the clock transition along the lattice direction to access the Lamb-Dicke regime.  After applying the spectroscopic pulse, the fluorescence from a resonant 398.9\,nm detection pulse ($<\,100\,\mu$s) was collected by a photomultiplier, thus measuring the number of atoms remaining in the ground state.  The timing of the spectroscopic and detection pulses was chosen such that most of the atoms held by the lattice were not lost, but atoms not trapped had time to accelerate out of the detection region. Lineshapes were generated by repeating this cycle and stepping the frequency of the spectroscopic laser.

Crucial to the lattice clock scheme is the ability to cancel out the large ac Stark perturbation on the clock transition due to the lattice.  This is accomplished by tuning the lattice wavelength such that the shifts on the ground and excited states are equal, leaving the clock frequency unperturbed.  This Stark-free wavelength had been theoretically estimated to be near 752\,nm \cite{Porsev04}.  The laser for the optical lattice consisted of a cw ring Ti:sapphire laser pumped at 532\,nm.  The Ti:sapphire laser was injection-locked by an external cavity diode laser to provide unidirectional and single frequency lasing with up to 2.2 W of output power \cite{Cummings02}.  This system was tunable from 750\,nm to 795\,nm, allowing a large range in which to search for the Stark-free lattice wavelength.  The lattice light was coupled using a polarization maintaining fiber to the MOT chamber area where it was collimated to a beam size of $\sim$3\,mm (1/$e^2$ radius). The beam was then focused using a 35\,cm focal length achromatic lens into the chamber and onto the atoms.  It was then retroreflected by a curved mirror after the chamber.  The resultant beam waist of $w_0\approx30\,\mu$m produced an optical lattice depth of tens of $\mu$K for 1\,W incident lattice power.

The yellow spectroscopic laser light at 578.42\,nm was provided by a highly stable dye laser that has been well described elsewhere \cite{Young99}. This laser has a linewidth of $<$\,1\,Hz with typical drift rates of several hundred hertz per day for its ultrastable reference cavity.  The laser and stabilization system were located in another laboratory, necessitating fiber delivery ($\sim200$\,m) with active noise cancellation \cite{Bergquist92, Longshen94}.  The spectroscopic light was scanned and chopped using an acousto-optic modulator (AOM).  After passing through the AOM, the spectroscopic light was combined with the lattice light using dichroic optics and focused with a common lens (see Fig. \ref{setup}), giving a beam waist approximately twice that of the lattice beam.  The retroreflecting mirror for the lattice was low-reflecting at the clock wavelength, ensuring very little standing wave structure of the probe light. 

\begin{figure}
\includegraphics[width=9cm]{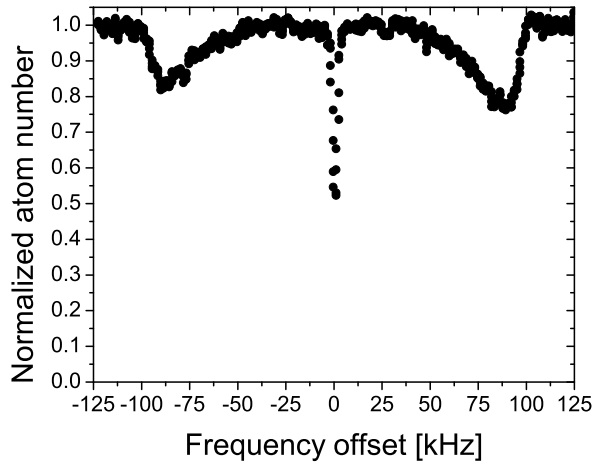}
\caption{Sideband spectrum of the magnetic field-induced clock transition of $^{174}$Yb atoms confined to an optical lattice ($|B|\approx$\,10\,mT).  The sideband spacing of $\omega_0/2\pi=90$\,kHz implies an average lattice depth of 1\,MHz ($\approx50\,\mu$K).  The mean vibrational quantum number can be found from the ratio of the lower to the upper sideband amplitude, which is $\left\langle n \right\rangle/(\left\langle n\right\rangle +1)$.  This yields a value of $\left\langle n\right\rangle=2.4$, which indicates an average lattice temperature of 15~$\mu$K.}

\label{sideband}
\end{figure}

An experimentally important parameter was the magnetic field-induced optical Rabi frequency \cite{Taichenachev05}.  From perturbation theory the $(6s6p)\,^3\!P_0$ clock state acquires a first order correction due to nearby states in the presence of the magnetic field.  The dominant mixing term occurs with the $(6s6p)\,^3\!P_1,m_J=0$ state, since all the other terms have either vanishing magnetic dipole matrix elements or are much further detuned.  The induced Rabi frequency is then given by $\Omega=\Omega_L\Omega_B/\Delta$, where $\Omega_L$ is the optical Rabi frequency for the 578.42\,nm clock laser on the $(6s^2)^1\!S_0\leftrightarrow(6s6p)\,^3\!P_1,m_J=0$ transition, $\Omega_B$ is the magnetic dipole matrix element $\langle ^3\!P_1,m_J=0|\vec{\mu}\cdot{\bf B}|^3\!P_0\rangle/\hbar$, and $\Delta$ is the splitting between the $^3\!P$ states (see Fig. \ref{setup}). For light polarization parallel to ${\bf B}$, $\Omega = \alpha\sqrt{I}|{\bf B}|$, where $I$ is the light field intensity.  The coefficient $\alpha$ contains all the atomic parameters, and for Yb its value is  $186$\,Hz/(T$\sqrt{\textrm{mW/cm$^2$}}$).  Estimated values of $\alpha$ for Sr, Ca, and Mg are tabulated in Ref. \cite{Taichenachev05}.  One finds that for experimentally reasonable fields significant Rabi frequencies can be generated (e.g., $|B|=2$\,mT, $I=10$\,mW/cm$^2$ gives $\Omega\approx$\,1\,Hz). 

To optimize the performance of the lattice apparatus and roughly locate the Stark-free wavelength of Yb, we first performed measurements on the more easily excited $^{173}$Yb clock transition.  Utilizing knowledge of the isotope shifts and the frequencies of the two odd $^{171,173}$Yb isotope resonances \cite{Hoyt05}, we were able to limit our search for the very narrow $^{174}$Yb resonance to $-550\pm5$\,MHz with respect to the $^{173}$Yb resonance.  To create the bias magnetic field needed for the even isotopes, we built a circuit to reverse the current direction in one of the MOT coils. This produced a Helmholtz-like configuration with a large uniform magnetic field.  With full magnetic field ($\sim$\,11\,mT) and optical power ($\sim$\,25\,W/cm$^2$) we could produce induced Rabi frequencies of $\sim$\,300\,Hz.  This greatly aided in the initial search for the resonance by accessing a highly power-broadened regime and also allowed us to investigate large values of the quadratic Zeeman shift.  Figure \ref{sideband} shows a spectrum taken at a lattice wavelength of 759.351\,nm with $|B|\approx10$\,mT. The narrow carrier in the center has asymmetric sidebands that have a spacing of $\omega_0/2\pi=90$\,kHz. This spacing corresponds to an average lattice depth of 1\,MHz ($\sim\,50\,\mu$K).  The mean vibrational quantum number, $\left\langle n\right\rangle=2.4$, can be determined from the asymmetry of the sidebands, which is equal to $\left\langle n \right\rangle/(\left\langle n\right\rangle +1)$ \cite{Wineland87}.  This asymmetry indicates an average temperature of 15\,$\mu$K for the atoms trapped in the lattice.

\begin{figure}
\includegraphics[width=9cm]{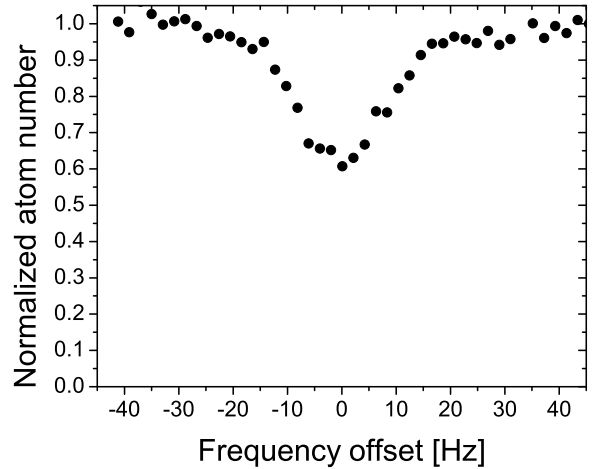}
\caption{Spectrum of the magnetic field-induced $^1\!S_0 \leftrightarrow\,^3\!P_0$ resonance with the lattice laser tuned close to the Stark-free wavelength.  A single 400\,ms measurement cycle consists of the two successive MOT stages, loading the lattice, applying a 64\,ms spectroscopic pulse ($|B|=1.29$\,mT, $I\approx280$\,mW/cm$^2$), then measuring the ground state population with a 398.9\,nm pulse. The frequency of the probe laser was stepped fifty times in 2\,Hz intervals for a total scan time of 20\,s.  A Gaussian fit to the line yields a FWHM of 20.3\,Hz.  The depletion of 40\,\% was achieved with single pulse excitation.}
\label{narrowline}
\end{figure}

Following the initial observation of the line, we determined the Stark-free wavelength by varying the lattice wavelength and observing the resulting frequency shifts and linewidths.  The large intensity variation of the lattice causes line broadening on the order of the differential ac Stark shift, which grows as the lattice is detuned from the Stark-free wavelength.  Monitoring the line broadening and the direction of the shift with increasing lattice intensity, we were able to determine the Stark-free wavelength for Yb to be $759.35\pm.02$\,nm.  This wavelength is located near an allowed two-photon resonance, $(6s6p)\,^3\!P_0\rightarrow (6s8p)\,^3\!P_0$ at 759.7\,nm. This may create a non-negligible differential light shift proportional to $I^2$ (ac hyperpolarizablity) that has to be evaluated in a high accuracy clock scheme \cite{Porsev04}.  

A spectrum of the carrier with the lattice tuned near the Stark-free wavelength is shown in Fig. \ref{narrowline}.  The 20\,Hz FWHM is close to the Fourier limited width due to the 64\,ms interrogation pulse ($\approx$14\,Hz). For pulse times less than $64$\,ms, we observed nominally Fourier limited lineshapes.  Increasing the pulse duration beyond 64\,ms, however, did not reduce the linewidth below 20\,Hz, indicating broadening mechanisms.  A possible mechanism is coupling of the transverse motion of the atoms in the 1D lattice through imperfect alignment and nonplanar wavefronts of the focused spectroscopic beam.  In addition, longitudinal vibrations of the lattice relative to the probe could also contribute. Another possibility is broadening from collisions between the 10 to 15 bosonic $^{174}$Yb atoms in each lattice site (density $\sim10^{10}$cm$^{-3}$).  Finally, a more careful determination of the Stark-free wavelength is necessary.  The 20\,pm ($\sim$\,10\,GHz) uncertainty could lead to an estimated shift of $\sim$\,180\,Hz with some broadening. 

The spectrum shown in Fig. \ref{narrowline} has a signal-to-noise ratio of approximately 10, which, with our present measurement cycle, should enable a measurement precision of 2\,Hz in one second.  With a clock frequency of 518\,THz, this present level of performance could support a fractional frequency instability of  $4\times10^{-15}$ for one second averaging time.  This is competitive with the performance of the best existing frequency standards.  Moreover, the use of normalized detection and modest improvements in the linewidth should yield improvement by an order of magnitude in the near future.  While estimates of the absolute frequency uncertainty of this clock system (e.g. collisional shifts) will require careful evaluations, uncertainties specific to this method could be held well below 10\,mHz for reasonable spectroscopic parameters \cite{Taichenachev05}.

To check that the excitation-enabling magnetic field does not lead to unmanageable clock frequency shifts, we measured the quadratic Zeeman shift coefficient, $\beta$.  Using the $m_J=\pm1$ splitting of the $^3P_1$ state to calibrate the magnetic field, we determined $\beta=-6.6\pm0.4$\,MHz/T$^2$, in agreement with the theoretical estimate of 6.2\,MHz/T$^2$ \cite{Taichenachev05}.  At the -1.29\,mT field used for the spectroscopy in Fig. \ref{narrowline} this gives a magnetic shift of only -11\,Hz.  A careful calibration yielding 0.5\,$\mu$T (5\,mG) uncertainty for a 1\,mT (10\,G) bias field would lead to a magnetic field-dependent uncertainty of 6\,mHz.

The benefits of using the even isotopes are considerable.  One of the largest frequency uncertainties of the Sr studies based on the odd isotopes \cite{Takamoto05, Ludlow05} was due to the linear Zeeman shift, which is further exacerbated by the complex ($I=9/2$) level structure.  In comparison, the even isotopes have no magnetic substructure and only quadratic dependence of the magnetic field, making the uncertainty due to the magnetic field with our method nearly negligible.  In addition, the scalar nature of the transition makes potential shifts due to the polarization of the lattice virtually nonexistent.  This will simplify the move to multidimensional lattices, which help to eliminate collisional shifts, and may enable better control of frequency shifts due to the hyperpolarizability.

In this Letter, we have demonstrated the first lattice-based spectroscopy with Yb atoms. This is also the first even-isotope study of the $^1\!S_0 \leftrightarrow\,^3\!P_0$ clock transition in an alkaline earth-like atom. We have observed linewidths as narrow as 20\,Hz FWHM for $^{174}$Yb atoms confined to a 1D optical lattice, yielding stability comparable with the best existing frequency standards.  This magnetic field-induced technique is simple to implement, applicable to many alkaline earth-like atoms, and allows the use of the metrologically superior even isotopes in optical lattice clocks.  The experimental simplification associated with this method should greatly accelerate the development of lattice-based clocks as future time standards.
 
We thank J.\ C.\ Bergquist for useful discussions and invaluable assistance with the clock laser, and Y.\ -J.\ Wang and E.\ A.\ Donley for their careful reading of the manuscript. We appreciate experimental help from O.\ Ovchinnikova.  C.\ W.\ Hoyt is grateful for support from the National Research Council.  A.\ V.\ Taichenachev and V.\ I.\ Yudin were supported by RFBR (04-02-16488,05-02-17086, 05-08-01389).


\end{document}